\documentclass[12pt]{article}
\usepackage{epsfig}
\newtheorem{axiom}{Axiom}
\newcommand{\bax}{\begin{axiom}}
\newcommand{\eax}{\end{axiom}}
\newcommand{\bra}[1]{\langle#1\vert}
\newcommand{\ket}[1]{\vert#1\rangle}

\newcommand{\braket}[2]{\langle#1\vert#2\rangle}
\newcommand{\ketbra}[2]{\vert#1\rangle\langle#2\vert}
\newcommand{\sandwich}[3]{\langle#1\vert#2\vert#3\rangle}
\newcommand{\Tr}{\hbox{Tr}}
\newcommand{\DC}{\!{\cal DC}\,}
\newcommand{\be}{\begin{equation}}
\newcommand{\ee}{\end{equation}}
\newcommand{\bea}{\begin{eqnarray}}
\newcommand{\eea}{\end{eqnarray}}
\newcommand{\bq}{\begin{quote}}
\newcommand{\eq}{\end{quote}}
\newcommand{\bd}{\begin{description}}
\newcommand{\ed}{\end{description}}
\newcommand{\ben}{\begin{enumerate}}
\newcommand{\een}{\end{enumerate}}
\newcommand{\bi}{\begin{itemize}}
\newcommand{\ei}{\end{itemize}}
\renewcommand{\d}{\partial}
\newcommand{\bA}{{\bf A}}

\newcommand{\bp}{{\bf p}}
\newcommand{\br}{{\bf r}}

\newcommand{\cA}{{\cal A}}
\newcommand{\cB}{{\cal B}}
\newcommand{\cC}{{\cal C}}

\newcommand{\cH}{{\cal H}}
\newcommand{\cL}{{\cal L}}
\newcommand{\cM}{{\cal M}}
\newcommand{\cN}{{\cal N}}

\newcommand{\cP}{{\cal P}}

\newcommand{\cS}{{\cal S}}

\newcommand{\cU}{{\cal U}}
\newcommand{\cV}{{\cal V}}

\newcommand{\hA}{{\bf\hat A}}
\newcommand{\hB}{{\bf\hat B}}
\newcommand{\hI}{{\bf\hat 1}}
\newcommand{\hH}{{\bf\hat H}}
\newcommand{\hM}{{\bf\hat M}}
\newcommand{\hN}{{\bf\hat N}}
\newcommand{\hO}{{\bf\hat O}}
\newcommand{\hP}{{\bf\hat P}}
\newcommand{\hU}{{\bf\hat U}}
\newcommand{\hW}{{\bf\hat W}}
\begin{document}
\setlength{\baselineskip}{16.5pt}
\title{Quantum mechanics explained}
\author{U. Mohrhoff\\
Sri Aurobindo International Centre of Education\\
Pondicherry 605002 India\\
\normalsize\tt ujm@auromail.net}
\date{}
\maketitle
\begin{abstract}
\noindent The physical motivation for the mathematical formalism of quantum mechanics is made clear and compelling by starting from an obvious fact---essentially, the stability of matter---and inquiring into its preconditions: what does it take to make this fact possible?

\vspace{6pt}\noindent {\it Keywords\/}: fuzziness; interpretation; probability; quantum mechanics; stability

\vspace{6pt}\noindent {\it PACS numbers\/}: 01.40.Gm; 03.65.-w; 03.65.Ca; 03.65.Ta
\setlength{\baselineskip}{14pt}
\end{abstract}
\pagebreak
\section{Introduction}
In the very first section of his brilliant Caltech lectures,~\cite{Feynmanteach} Feynman raises a question of concern to every physics teacher: ``Should we teach the {\it correct\/} but unfamiliar law with its strange and difficult conceptual ideas~\dots? Or should we first teach the simple~\dots law, which is only approximate, but does not involve such difficult ideas? The first is more exciting, more wonderful, and more fun, but the second is easier to get at first, and is a first step to a real understanding of the second idea.'' With all due respect to one of the greatest physicists of the 20th Century, I cannot bring myself to agree. How can the second approach be a step to a real understanding of the correct law if ``{\it philosophically we are completely wrong\/} with the approximate law,'' as Feynman himself emphasizes in the immediately preceding paragraph? To first teach laws that are completely wrong philosophically cannot but impart a conceptual framework that eventually stands in the way of understanding the correct laws. The damage done by imparting philosophically wrong ideas to young students is not easily repaired. 

Feynman himself has demonstrated that a theory as complex as quantum electrodynamics can be explained to a general audience in just four lectures~\cite{QED}. He even told his audience that ``[b]y explaining quantum electrodynamics to you in terms of what we are \textit{really doing} [original emphasis], I hope you will be able to understand it better than do some of the students.'' If it is possible, in four lectures, to make a lay audience understand quantum electrodynamics better than some students of physics, then there has to be something wrong with the conventional methods of teaching physics.

What we are \textit{really doing} is calculating the probabilities of measurement outcomes. The frank admission that quantum physics is simply a collection of probability algorithms makes a meaningful introduction to quantum physics almost as easy as teaching classical physics. I teach quantum mechanics to undergraduates and even high school students (grades 11--12) at the Sri Aurobindo International Centre of Education (SAICE) in Pondicherry, India, introducing the subject with a statement of two simple premises and two simple rules. These suffice to discuss key quantum phenomena in considerable depth and without much further mathematical ado. They are discussed in Secs.~11 and 12 below.

I believe that one of the reasons why we find it so hard to beat sense into quantum physics is that it uses much the same concepts as its classical counterpart (e.g., position, time, energy, momentum, angular momentum, mass, charge), and that these concepts are rarely adequately revised when students advance from classical to quantum. Instead of examining how and why the classical concepts emerge in those conditions in which quantum physics degenerates into classical physics, we ``quantize'' the classical concepts---a mathematical procedure supported by little if any physical insight. The frank admission that quantum mechanics is simply a set of tools for assigning probabilities to measurement outcomes makes it easy to see that the concepts of classical physics, being rooted in the mathematical features of a probability algorithm, have \textit{no kind of ontological} basis. The world isn't made of positions, or energy, or mass, or any combination of such things.

But is it true that quantum physics is simply a collection of probability algorithms? In this article I present 
a derivation of the theory's formalism to which my students respond well. It is (weakly) teleological in that it takes an obvious fact---essentially, the stability of matter---and inquires into its preconditions: what does it take to make it possible? While this approach lacks the rigor of purely axiomatic approaches (e.g., Refs. \cite{Mackey}--\cite{Pitowsky89})---it is, after all, intended for high school students and undergraduates---it makes the standard axioms of quantum mechanics seem transparent and well-motivated. In addition it should make it obvious that the mathematical formalism of quantum physics is indeed nothing but a probability algorithm. (I shall skip over definitions and explanations that readers of this journal would in all likelihood find excessive.)

Section 2 introduces the classical probability algorithm, which only assigns trivial probabilities (in the particular sense spelled out there). Section~3 explains why we need a probability algorithm that can accommodate nontrivial probabilities. Section~4 initiates a straightforward transition from classical to quantum, by upgrading from a probability algorithm that is equivalent to a point to a probability algorithm that is equivalent to a line. Three axioms uniquely determine the structure of this algorithm. They are introduced, respectively, in Secs. 5, 6, and 7. At the heart of the resulting algorithm is the trace rule, according to which the probabilities of the possible outcomes of measurements are determined by a density operator, whose properties are discussed in Sec.~8. Section~9 establishes how the probabilities of possible measurement outcomes depend (via the density operator) on actual measurement outcomes, and Sec.~10 explains how they depend on the times of measurements. In Sec.~13 the two rules derived in Secs. 11 and 12 are used to introduce the propagator for a free and stable scalar particle, to explain why quantum mechanics needs complex numbers, and to trace the meaning of ``mass'' to its algorithmic origins. Section~14 does the same for ``energy'' and ``momentum'' and shows how the quantum-mechanical probability algorithms make room for electromagnetic effects. The penultimate section introduces the wave function and recalls Feynman's derivation of the Schr\"odinger equation. The final section lists the typical postulates of standard axiomatizations of quantum mechanics and points out the fallacies involved, owing to which these postulates seem anything but clear and compelling. Once they are removed, however, these postulates turn out to define the same theoretical structure as that derived in this article.

\section{The classical probability algorithm}
\label{SecClPA}The state of a classical system at any particular time is given by the values of a definite and unchanging number of coordinates and an equal number of momenta. The state of a classical system with $n$ degrees of freedom can therefore be represented by a point~$\cP$ in a $2n$-dimensional phase space~$\cS$, and the positive outcome of every possible elementary test can be represented by a subset~$\cU$ of~$\cS$. (An elementary test has exactly two possible outcomes. It assigns a truth value---``true'' or ``false''---to a proposition of the form ``system~$S$ has property~$P$.'') The probability of obtaining~$\cU$ is 1 if $\cP\in\cU$ and 0 if $\cP\not\in\cU$. This probability algorithm is trivial in the sense that it only assigns the trivial probabilities 0 or~1. Because it is trivial, it can be thought of as a state in the classical sense of the word: a collection of possessed properties. We are therefore free to believe that the probability of finding the property represented by~$\cU$ is~1 \textit{because} the system has this property.

\section{Why nontrivial probabilities?}
\label{SECwhyprobas}One spectacular failures of classical physics was its inability to account for the stability of matter. ``Ordinary'' material objects
\bi
\item have spatial extent (they ``occupy space''),
\item are composed of a (large but) finite number of objects without spatial extent (particles that do not ``occupy space''),
\item and are stable: they neither explode nor collapse as soon as they are created.
\ei
Ordinary objects occupy as much space as they do because atoms and molecules occupy as much space as {\it they\/} do. So how is it that a hydrogen atom in its ground state occupies a space roughly one tenth of a nanometer across? Thanks to quantum mechanics, we now understand that the stability of ordinary objects rests on the {\it fuzziness\/} of the relative positions and momenta of their constituents. It is rather unfortunate that Heisenberg's term \textit{Unsch\"arfe}---as in \textit{Unsch\"arfe-Prinzip} or \textit{Unsch\"arfe-Relation} has come to be translated as ``uncertainty.'' The primary dictionary definition of \textit{Unsch\"arfe} is ``fuzziness'' (followed by ``haziness'', ``diffusion'', ``poor definition'', ``blur'', and ``blurring''). What ``fluffs out'' matter is not our \textit{subjective} uncertainty about the positions of atomic electrons relative to atomic nuclei but the \textit{objective} fuzziness of these positions. (Lieb~\cite{Lieb} has obtained the following result: if we assume that the attraction between electrons and protons varies as~$1/r^2$, and that the number of protons in a nucleus has an upper limit, then the stability of ordinary material objects requires both the Heisenberg uncertainty principle and the Pauli exclusion principle.)

What, then, is the proper---mathematically rigorous and philosophically sound---way to define and quantify a fuzzy observable? I submit that it is to assign nontrivial probabilities---probabilities anywhere between 0 and~1---to the possible outcomes of a measurement of this observable. To be precise, the proper way of dealing with a fuzzy observable is to make \textit{counterfactual} probability assignments. For obvious reasons, a probability distribution over the possible outcomes of a position measurement that could be made at the time~$t$, determined by the outcome of an earlier measurement, can represent the fuzzy position of a particle at the time~$t$ only if no measurement is actually made.

The invariable reference to ``measurement'' in standard axiomatizations of quantum mechanics was famously criticized by Bell: ``To restrict quantum mechanics to be exclusively about piddling laboratory operations is to betray the great enterprise'' \cite{Bell}. Yet neither can the unperformed measurements that play a key role in the quantitative description of fuzzy observables be characterized as ``piddling laboratory operations,'' nor is the occurrence of measurements restricted to physics laboratories. \textit{Any} event from which either the truth or the falsity of a proposition of the form ``system~S has property~P'' can be inferred, qualifies as a measurement.

\section{Upgrading from classical to quantum}
\label{SecUpgrade}The classical algorithm of Sec.~\ref{SecClPA} cannot accommodate the nontrivial probabilities that we need for the purpose of defining and quantifying a fuzzy observable. Nor can the probability algorithms of classical \textit{statistical} physics---distributions over a phase space---be used for this purpose. The reason this is so is that the nontrivial probabilities of classical physics are  \textit{subjective}. They are ignorance probabilities, which enter the picture when relevant facts are ignored, and which disappear, or degenerate into trivial probabilities, when all relevant facts are taken into account. The so-called ``uncertainty principle,'' on the other hand, guarantees that quantum-mechanical probabilities cannot be made to disappear. Quantum-mechanical probabilities are therefore \textit{objective}.

My characterization of quantum-mechanical probabilities as objective should not be confused with the frequentist's definition of probability and his/her characterization of relative frequencies as ``objective probabilities.'' Classical statistical mechanics is objective in much the same sense in which relative frequencies are objective. It is at the same time subjective in the sense that it ignores objectively existing features of the systems it studies. If the properties possessed by the members of a classical ensemble at a given time were known with absolute precision, then it would be possible in principle to assign, for every later time, a trivial probability to every subset of the ensemble's space of states. Quantum-mechanical probabilities are the only probabilities we know that are not subjective in this sense. As Mermin has pointed out, ``quantum mechanics is the first example in human experience where probabilities play an essential role even when there is nothing to be ignorant about''~\cite{Mermin}. (There is one classical single-case probability that might be considered objective: the propensity of a nucleus to decay within a given interval of time if radioactive decay is treated as a classical stochastic process. But this treatment of radioactive decay is obviously nothing but a classical mock-up of the correct, quantum-mechanical treatment.)

There is another sense in which quantum-mechanical probabilities are objective. They are objective not only because they serve to define and quantify an objective fuzziness but also because the algorithms that we use to calculate them form part of the fundamental theoretical framework of contemporary physics. Quantum-mechanical probabilities are \textit{conditional} probabilities~\cite{Primas}. We use them to assign probabilities to possible measurement outcomes \textit{on the basis of actual outcomes}. In other words, the so-called ``dynamical'' laws of quantum physics encapsulate \textit{correlations} (between measurement outcomes). Quantum-mechanical probabilities therefore depend (i)~on objective correlation laws and (ii)~on objective measurement outcomes. This rules out an epistemic interpretation of probability in the context of quantum theory.

Arguably the most straightforward way to make room for nontrivial objective probabilities is to upgrade from a 0-dimensional point to a 1-dimensional line. Instead of representing our probability algorithm by a point in a phase space, we represent it by a 1-dimensional subspace of a vector space~$\cV$ (a ``ray''). And instead of representing measurement outcomes by subsets of a phase space, we represent them by subspaces of this vector space. (In what follows "subspace" will be short for "closed subspace.") A 1-dimensional subspace~$\cL$ can be contained in a subspace~$\cU$, it can be orthogonal to~$\cU$ ($\cL\perp\cU$), but now there is a third possibility, and this is what makes room for nontrivial probabilities. $\cL$~assigns probability~1 to outcomes represented by subspaces containing~$\cL$, it assigns probability~0 to outcomes represented by subspaces orthogonal to~$\cL$, and it assigns probabilities greater than~0 and less than~1 to measurement outcomes represented by subspaces that neither contain nor are orthogonal to~$\cL$. (The virtual inevitability of this ``upgrade'' was demonstrated by Jauch~\cite{Jauch}.)

\section{The first postulate}
Following Dirac, we denote a vector by the symbol $\ket v$, and if the vectors $\{\ket{a_i}|i=1,\dots,n\}$ form an orthonormal basis (ONB), we shall denote the components of $\ket v$ with respect to this basis by $\braket{a_i}v$. Thus,
\be
\ket v=\sum_i \ket{a_i}\braket{a_i}v.
\ee
Will a real vector space do or are complex components needed? Pending the final decision, we shall assume that $\cV$ is complex. (If a real vector space turned out to be sufficient, it would be easy to appropriately constrain our conclusions.) We must therefore keep in mind that $\braket ba=\braket ab^*$. The advantage of Dirac's notation is that it consolidates two different ways of thinking about the expression $\ket{a_i}\braket{a_i}v$:
\be
\ketbra{a_i}{a_i}\quad\ket v\quad = \quad\ket{a_i}\quad\braket{a_i}v.
\label{EQDirac}
\ee
$\bra{a_i}$ is a machine called \textit{covector}, which accepts a vector $\ket v$ and returns a complex number, the \textit{component} $\braket{a_i}v$ of $\ket v$ with respect to the basis vector $\ket{a_i}$. $\ketbra{a_i}{a_i}$ is a machine called \textit{projector}, which accepts a vector $\ket v$ and returns another vector, the \textit{projection} of~$\ket v$ into the 1-dimensional subspace containing $\ket{a_i}$. Think of the left-hand side of Eq.~(\ref{EQDirac}) as what we do---projecting $\ket v$ into the 1-dimensional subspace containing $\ket{a_i}$---and of the right hand side as what we get as a result: the basis vector $\ket{a_i}$ multiplied by the component of~$\ket v$ with respect to~$\ket{a_i}$.

There is an obvious one-to-one correspondence between subspaces and projectors. If $\cU$ is an $m$-dimensional subspace of~$\cV$, then there is a set of mutually orthogonal unit vectors $\{\ket{b_k}|k=1,\dots,m\}$ such that (i)~every vector in $\cU$ can be written as a unique linear combination of this set of vectors, and (ii)~the operator
\be
\hP=\sum_{k=1}^m \ketbra{b_k}{b_k}
\label{EQprojector}
\ee
projects into $\cU$. Hence instead of representing measurement outcomes by subspaces, we may represent them by the corresponding projectors. Our first axiom thus reads,
\bax
Measurement outcomes are represented by the projectors of a vector space.
\eax
\section{The second postulate}
\label{SecAxiom2}Unsurprisingly, my course begins with a discussion of ``the most beautiful experiment'' \cite{Crease}---the two-slit experiment with electrons~\cite{FLS}--\cite{Tonomura}. This is followed by a discussion of the ESW experiment (named after Englert, Scully, and Walther~\cite{SEW}--\cite{Mohrhoff99}), in which the experimenters have a choice between two incompatible measurements. The first measurement answers the question: through which slit did the atom go? The second measurement asks: how did the atom go through both slits---in phase or out of phase? The first question implicitly assumes that the atom went through a single slit. The second question involves the assumption that the atom went through both slits. The reason why these measurements are physically incompatible---i.e., incapable of being simultaneously made---is that the questions they are designed to answer are \textit{logically} incompatible: if one makes sense, the other does not.
\begin{figure}[t]
\begin{center}
\epsfig{file=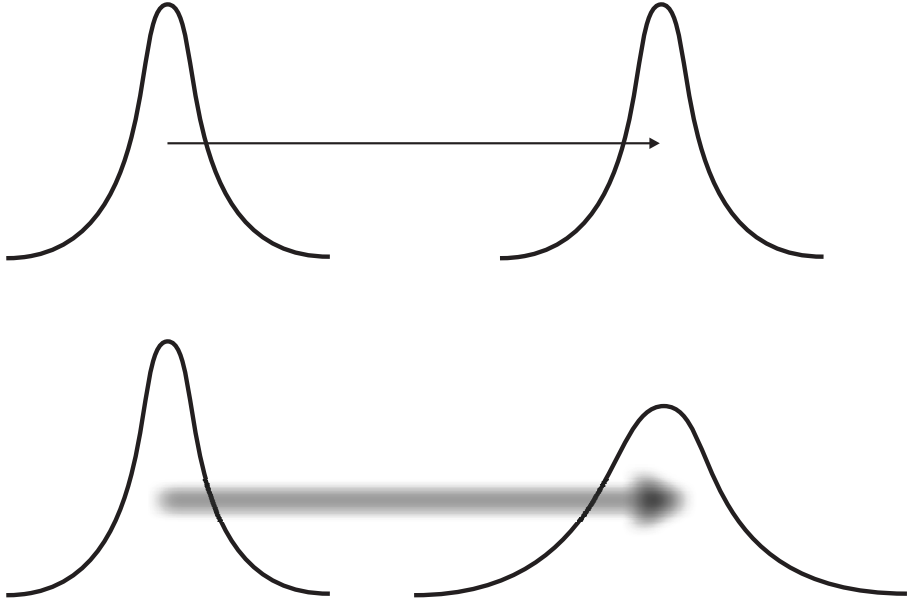,width=12cm}
\end{center}
\caption{A fuzzy position, represented as a probability distribution, at two different times. Above: if an object with this position moves with an exact momentum, then it moves by an exact distance. Below: if the same object moves with a fuzzy momentum, then it moves by a fuzzy distance; as a result, its position get fuzzier.}
\label{FIGuncty}
\end{figure}

Why does Nature confront us with incompatible measurements? To find out, imagine an object~$O$, composed of one positive and one negative charge, whose sole internal relative position is fuzzy. The standard deviation $\Delta r$ of the probability distribution associated with the radial component~$r$ of this relative position provides a handy measure of the fuzziness. If the electrostatic attraction between the two charges were the only force at work, it would cause a decrease in $\Delta r$, and $O$ would collapse as a result. What is needed for the stability of~$O$ is an effective repulsion that counterbalances the electrostatic attraction. Arguably the simplest and most direct way of obtaining this repulsion (considering that we already have a fuzzy position) is to let the conjugate momentum be fuzzy, too. Figure~\ref{FIGuncty} illustrates the fact that a fuzzy momentum causes a fuzzy position to grow fuzzier.

Even the fuzziness of both the relative position \textit{and} its conjugate momentum, however, are not sufficient for the existence of a stable equilibrium between the two forces or tendencies. If the mean distance between the two charges decreases, their electrostatic attraction increases. If there is to be a stable equilibrium, the effective repulsion, too, must increase. We will therefore not be surprised to find that a decrease in $\Delta r$ implies an increase in the fuzziness $\Delta p$ of the corresponding momentum, and vice versa. But this means that the probability distributions for a position and its conjugate momentum cannot simultaneously be dispersion-free: the product $\Delta r\,\Delta p$ will have a positive lower limit. It is therefore impossible to simultaneously measure both $r$ and~$p$ with arbitrary precision.

To arrive at a formal definition of compatibility, we consider two elementary tests. We begin with a situation in which a particular outcome of the second test (say, the positive outcome~$\cN$) is implied by an outcome of the first (say, the positive outcome~$\cM$). If we think of $\cM$ and~$\cN$ as outcomes, this means that $\cM\Rightarrow\cN$. If we think of them as subspaces, it means that $\cM\subset\cN$. It is left as an exercise to show that if the subspace $\cA$ represents the positive outcome of an elementary test, then the orthocomplement $\cA_\perp$ of~$\cA$ represents the negative outcome.

There are three possible combinations of outcomes: (i)~$\cM$ and~$\cN$, (ii)~$\cM_\perp$ and~$\cN$, and (iii)~$\cM_\perp$ and~$\cN_\perp$. We define compatibility as requiring that there be three 1-dimensional subspaces (``lines'') $\cL_i$ such that
\ben
\item $\cL_1\subseteq\cM\cap\cN$: $\cL_1$ assigns probability~1 to both $\cM$ and~$\cN$,
\item $\cL_2\subseteq\cM_\perp\cap\cN$: $\cL_2$ assigns probability~1 to both $\cM_\perp$ and~$\cN$,
\item $\cL_3\subseteq\cM_\perp\cap\cN_\perp$: $\cL_3$ assigns probability~1 to both $\cM_\perp$ and $\cN_\perp$.
\een
($\cA\cap\cB$ is the set-theoretic intersection of two subspaces, which is itself a subspace.) But if $\cM$ is a possible outcome and $\cM\subset\cN$ (which implies $\cN_\perp\subset\cM_\perp$), then $\cL_1$ exists. If $\cN_\perp$ is a possible outcome and $\cN_\perp\subset\cM_\perp$, then $\cL_3$ exists. And if $\cM\subset\cN$, then $\cN$ also contains a line that is orthogonal to~$\cM$, so that $\cL_2$ exists.

We now turn to the situation in which neither outcome of the first test implies either outcome of the second test. Compatibility then requires that there exists, in addition to the above three lines, a 1-dimensional subspace $\cL_4$ such that
\ben
\setcounter{enumi}{3}
\item $\cL_4\subseteq\cM\cap\cN_\perp$: $\cL_4$ assigns probability~1 to both $\cM$ and~$\cN_\perp$.
\een
None of the four intersections $\cM\cap\cN$, $\cM\cap\cN_\perp$, $\cM_\perp\cap\cN$, $\cM_\perp\cap\cN_\perp$ can now equal the 0-dimensional subspace~$\emptyset$ containing only the null vector. Since in addition they are mutually orthogonal, one can find a set of mutually orthogonal unit vectors $B=\{\ket{a_i}|i=1,\dots,n\}$ such that those with $i=1,\dots,j$ span $\cM\cap\cN$, those with $i=j+1,\dots,k$ span $\cM\cap\cN_\perp$, those with $i=k+1,\dots,m$ span $\cM_\perp\cap\cN$, and those with $i=m+1,\dots,n$ span $\cM_\perp\cap\cN_\perp$. To see that $B$ is, in fact, an ONB, we assume the contrary. We assume, in other words, that there is a 1-dimensional subspace~$\cL$ that is orthogonal to all four intersections or, what comes to the same, that there is a probability algorithm that assigns probability~0 to all possible combinations of outcomes of two compatible elementary tests. But this is a \textit{reductio ad absurdum} of our assumption. It follows that
\be
(\cM\cap\cN)\cup(\cM\cap\cN_\perp)\cup(\cM_\perp\cap\cN)\cup(\cM_\perp\cap\cN_\perp)=\cV.
\label{EQMNcupcap}
\ee
The \textit{span} $\cA\cup\cB$ of $\cA$ and~$\cB$---not to be confused with the set-theoretic union---is the smallest subspace that contains both $\cA$ and~$\cB$.
\begin{figure}[t]
\begin{center}
\epsfig{file=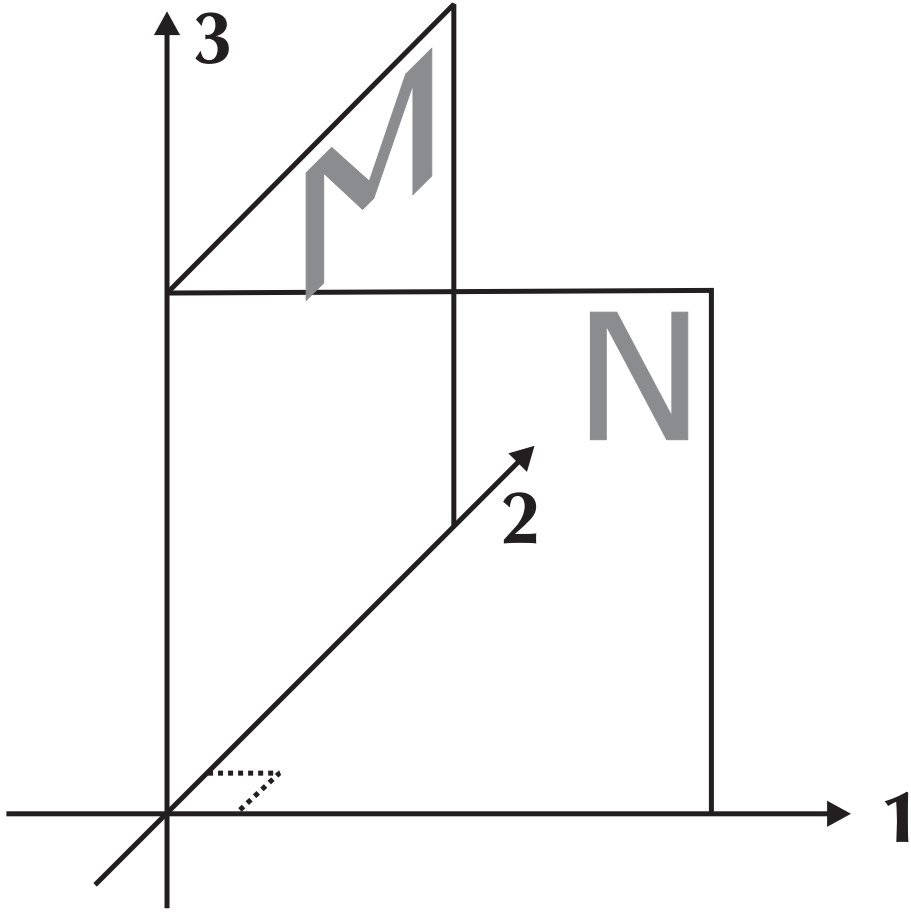,width=12cm}
\end{center}
\caption{Two subspaces with commuting projectors.}
\label{FIGcom}
\end{figure}

\begin{figure}[t]
\begin{center}
\epsfig{file=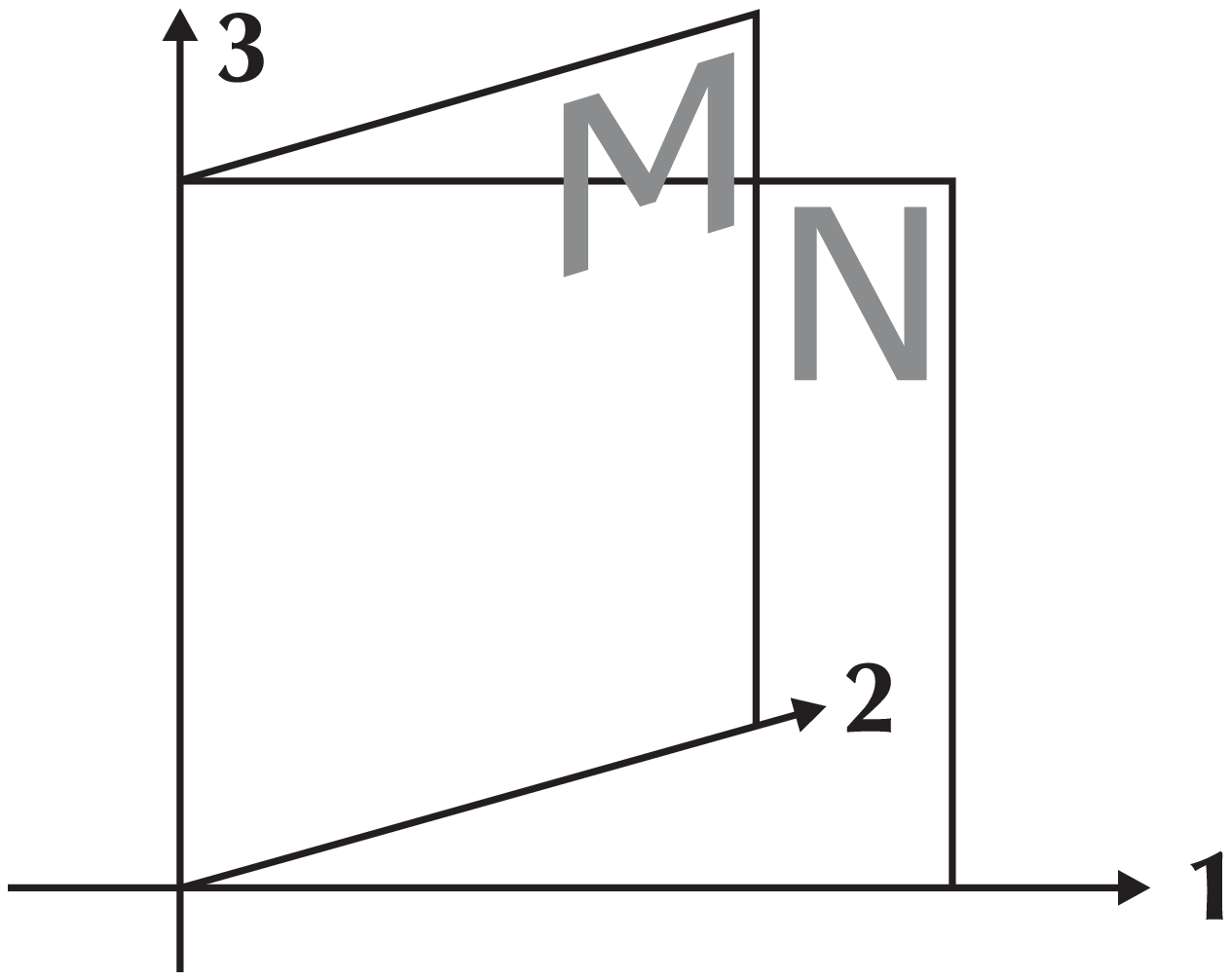,width=12cm}
\end{center}
\caption{Two subspaces with noncommuting projectors.}
\label{FIGnoncom}
\end{figure}

\begin{figure}[t]
\begin{center}
\epsfig{file=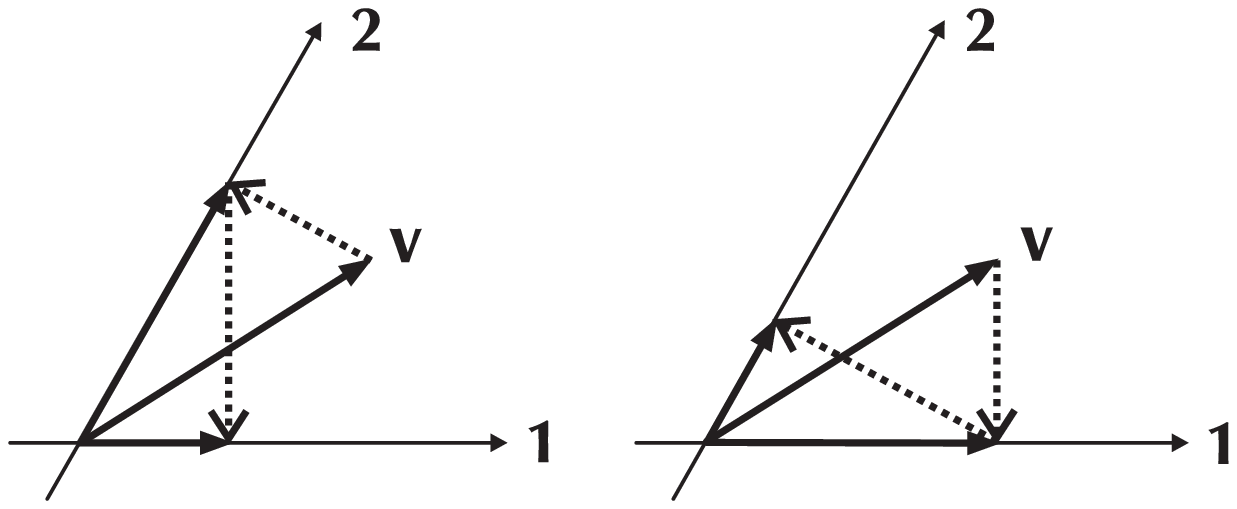,width=12cm}
\end{center}
\caption{The situation illustrated in Fig. \ref{FIGnoncom} seen from above, looking down the 3-axis. Left: first $\ket v$ is projected into~$\cM$ (the subspace containing~$\ket2$), then the resulting vector is projected into~$\cN$ (the subspace containing~$\ket1$). Right: first $\ket v$ is projected into~$\cN$, then the resulting vector is projected into~$\cM$.}
\label{FIGncvecs}
\end{figure}

Let us translate this results into the language of projectors. It is readily seen that two projectors $\hM$ and $\hN$ commute if and only if there is an ONB $\{\ket{a_i}|i=1,\dots,n\}$ such that
\be
\hM=\sum_{\hbox{{\scriptsize some} }i}\ketbra{a_i}{a_i}\qquad\hbox{and}\qquad
\hN=\sum_{\hbox{{\scriptsize some} }k}\ketbra{a_k}{a_k}.
\label{commonbasis}
\ee
As an illustration we consider the following example:
\be
\hM=\ketbra22+\ketbra33,\qquad \hN=\ketbra11+\ketbra33.
\ee
If the vectors $\ket1,\ket2,\ket3$ belong to the same ONB (Fig.~\ref{FIGcom}), commutability follows from the orthonormality relations $\braket{a_i}{a_k}=\delta_{ik}$. If at least one pair of vectors (say, $\ket1$ and $\ket2$) are not orthogonal, then
\be
\hM\,\hN=\ket2\braket21\bra1+\ketbra33,\qquad \hN\,\hM=\ket1\braket12\bra2+\ketbra33.
\ee
This is to say that $\hM\,\hN\neq\hN\,\hM$ (Fig.~\ref{FIGnoncom}). Figure~\ref{FIGncvecs} illustrates the respective actions of $\hN\,\hM$ and $\hM\,\hN$ on a vector~$\ket v$.

\begin{figure}[t]
\begin{center}
\epsfig{file=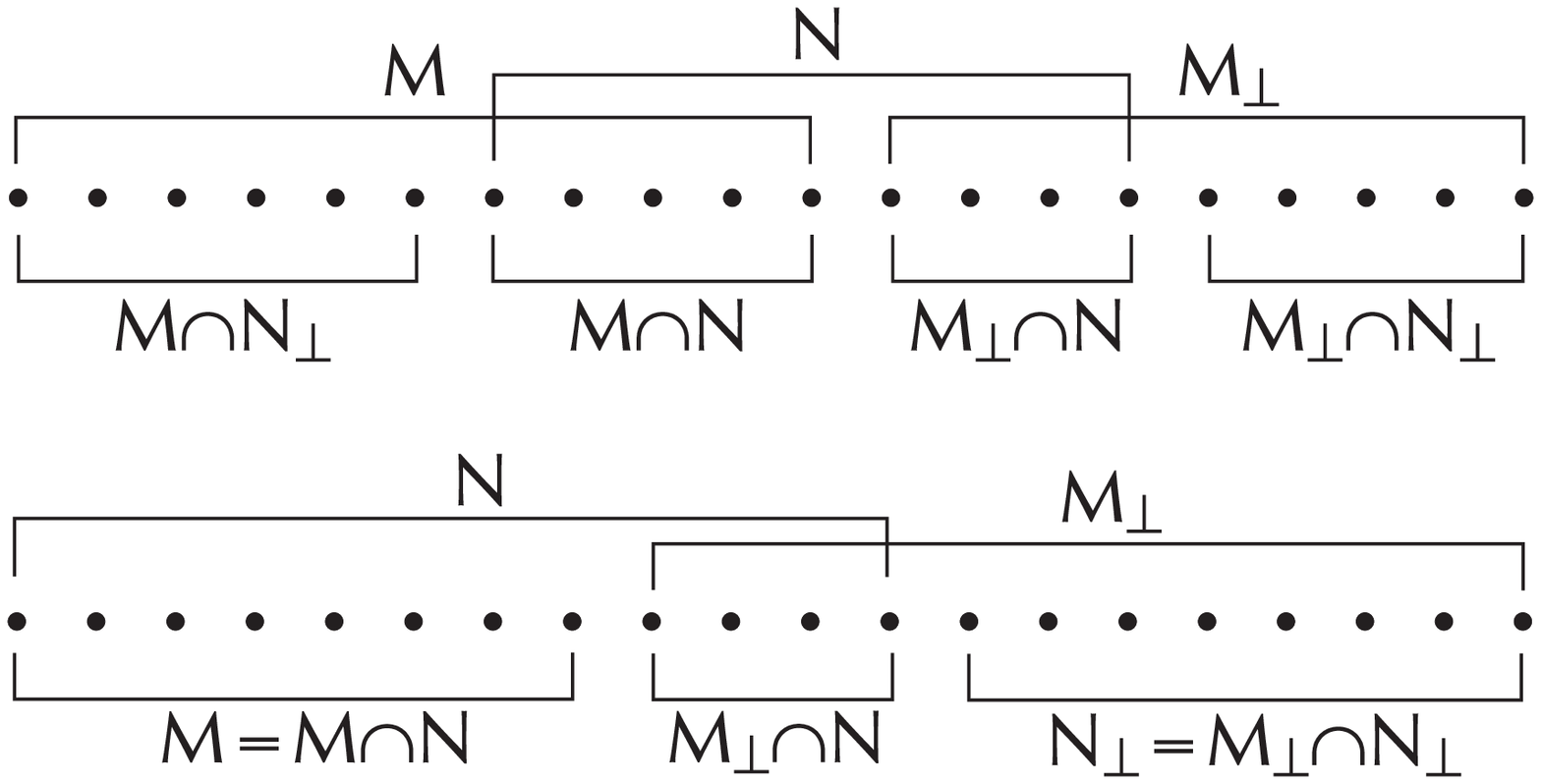,width=12cm}
\end{center}
\caption{The relations that hold between two subspaces $\cM$, $\cN$, their orthocomplements, and their intersections if the corresponding projectors can be written in terms of a common basis, as in~(\ref{commonbasis}). The dots represent the basis vectors. The horizontal bracket associated with a subspace groups the 1-dimensional projectors whose sum equals the projector onto this subspace. Above: the general case. Below: an outcome of one test implies an outcome of the other test (here $\cM\subset\cN$).}
\label{FIGcompro}
\end{figure}

It is also readily seen that Eq.~(\ref{EQMNcupcap}) holds if the projectors corresponding to the subspaces on the left-hand side of this relation can be written in terms of a common basis (Fig.~\ref{FIGcompro}), and that it does not hold if this is not the case. For the subspaces of Fig.~\ref{FIGnoncom}, for example, instead of Eq.~(\ref{EQMNcupcap}) we have that
\be
(\cM\cap\cN)\cup\emptyset\cup\emptyset\cup\emptyset=\cM\cap\cN\neq\cV.
\ee
Our second axiom therefore reads:
\bax
The outcomes of compatible elementary tests correspond to commuting projectors.
\label{AXcomcom}
\eax

\section{The third postulate}
Given two measurement outcomes represented by the subspaces $\cA$ and $\cB$, what measurement outcome is represented by the span of these subspaces, $\cA\cup\cB$?  Let $p(\cA)$ be the probability of obtaining the outcome represented by~$\cA$. Since a 1-dimensional subspace (or line) contained in either $\cA$ or~$\cB$ is contained in $\cA\cup\cB$, we have that
\be
\bigl[p(\cA)=1\;\hbox{or}\;p(\cB)=1\bigr]\;\Longrightarrow\;p(\cA\cup\cB)=1.
\ee
Since a line orthogonal to both $\cA$ and~$\cB$ is orthogonal to $\cA\cup\cB$, we also have that
\be
\bigl[p(\cA)=0\;\hbox{and}\;p(\cB)=0\bigr]\;\Longrightarrow\;p(\cA\cup\cB)=0.
\ee
This is what we expect if $\cA$ and $\cB$ represent two disjoint intervals $A$ and~$B$ in the range of a continuous observable~$O$. Note, however, that a 1-dimensional subspace can be in $\cA\cup\cB$ without being contained in either~$\cA$ or~$\cB$. In other words, obtaining the outcome $\cA\cup\cB$ can be certain even if neither obtaining~$\cA$ nor obtaining~$\cB$ is certain. The outcome $\cA\cup\cB$ therefore does not indicate that the value of~$O$ is \textit{either} $A$ \textit{or}~$B$ (let alone a number in either interval) but only that the value of $O$ is $A\cup B$. The possible outcomes of a continuous observable are the subsets of its range.

We now consider two measurements. The first, $M_1$, has three possible outcomes: $\cA$, $\cB$, and~$\cC$. The second, $M_2$, has two: $\cA\cup\cB$ and~$\cC$. Because the probabilities of all possible outcomes add up to~1, we have
\be
p(\cA)+p(\cB)+p(\cC)=1 \quad\hbox{as well as}\quad p(\cA\cup\cB)+p(\cC)=1.
\ee
It follows that
\be
p(\cA\cup\cB)=p(\cA)+p(\cB).
\label{EQaddi}
\ee
Or does it? Both measurements test for the possession of~$\cC$, but whereas $M_1$~makes two additional tests, $M_2$~only makes one. What gives us the right to demand \textit{non-contextuality}, i.e., that the probability of a test outcome be independent of whichever other tests are simultaneously made? We are looking for a probability algorithm that is capable of accommodating objective nontrivial probabilities. If requiring common sense in the form of noncontextuality helps us to pin down this algorithm, we go for it. By hindsight we know that Nature concurs. (Contextuality turns out to be an inescapable feature of situations where probabilities are assigned either on the basis of past \textit{and} future outcomes~\cite{Mohrhoff01} or to outcomes of measurements performed on ``entangled'' systems~\cite{PeresCh7}--\cite{Clifton}.) 

Let us once again translate our findings into the language of projectors. If $\cA$ and~$\cB$ are different outcomes of the same measurement, then the corresponding projectors $\hA$ and~$\hB$ are orthogonal and their sum $\hA+\hB$ is the projector corresponding to the outcome $\cA\cup\cB$. Our third axiom therefore reads:
\bax
If $\hA$ and~$\hB$ are orthogonal projectors, then the probability of the outcome represented by $\hA+\hB$ is the sum of the probabilities of the outcomes represented by $\hA$ and~$\hB$, respectively.
\eax
\section{The trace rule}
Axioms 1--3 are sufficient \cite{Peres190} to prove an important theorem due to A.~M. Gleason \cite{Gleason}--\cite{Cookeetal}, which holds for vector spaces with at least three dimensions. (More recently the validity of the theorem has been established for 2-dimensional vector spaces as well~\cite{Fuchs2001}--\cite{Cavesetal}, by generalizing from projector valued measures to positive operator valued measures~\cite{Peres9-5}.)
Gleason's theorem states that the probability of obtaining the outcome represented by the projector~$\hP$ is given by
\be
p(\hP)=\Tr(\hW\hP),
\label{EQtracerule}
\ee
where $\hW$ is a unique operator, known as {\it density operator\/}, which has the following properties: 
\bi
\item $\hW$ is \textit{linear}: $\hW\Bigl(\alpha\ket a+\beta\ket b\Bigr)=\alpha\,\hW\ket a+ \beta\,\hW\ket b$.
\item $\hW$ is \textit{self-adjoint}: $\sandwich a\hW b={\sandwich b\hW a}^*$.
\item $\hW$ is \textit{positive}: $\sandwich a\hW a\geq0$. 
\item $\Tr(\hW)=1$.
\item $\hW^2\leq\hW$.
\ei
Let us try to understand why $\hW$ has these properties.  If $\hP$ is a 1-dimensional projector $\ketbra vv$, then the trace rule~(\ref{EQtracerule}) reduces to $p(\hP)=\sandwich v\hW v$. The self-adjointness of $\hW$ ensures that the probability $\sandwich v\hW v$ is a real number, and the positivity of $\hW$ ensures that $\sandwich v\hW v$ does not come out negative.

The outcomes of a maximal test (otherwise known as ``complete measurement'') are represented by a complete system of (mutually orthogonal) 1-dimensional projectors $\{\ketbra{a_k}{a_k}|k=1,\dots,n\}$. $\Tr(\hW)=1$ ensures that the probabilities $\sandwich{a_i}\hW{a_i}$ of these outcomes add up to~1, and together with the positivity of~$\hW$, it ensures that no probability comes out greater than~1.

If the density operator is idempotent ($\hW^2=\hW$), then it is a 1-dimensional projector $\ketbra ww$ and it is called a \textit{pure state}; $\ket w$~is the corresponding \textit{state vector} (which is unique up to a phase factor), and the trace rule simplifies to $p(\hP)=\sandwich w{\hP}w$. (Since we began by upgrading from a point in a phase space to a line in a vector space, we are not surprised to find that a pure state projects into a 1-dimensional subspace.) If in addition $\hP=\ketbra vv$, the trace rule boils down to the familiar Born rule,
\be
p(\hP)=\braket wv\braket vw=|\braket vw|^2.
\label{EQBorn}
\ee
If $\hW^2<\hW$ then $\hW$ is called a \textit{mixed state}. What shall we make of this possibility? Since $\hW$ is self-adjoint, there exists an ONB of eigenvectors $\ket{w_k}$ of $\hW$ with real eigenvalues~$\lambda_k$ such that
\be
\hW=\sum_{k=1}^n\lambda_k\,\ketbra{w_k}{w_k}.
\ee
$\hW^2<\hW$ is therefore equivalent to
\be
\sum_k\lambda_k^2\,\ketbra{w_k}{w_k}<\sum_k\lambda_k\,\ketbra{w_k}{w_k},
\label{EQWspec}
\ee
from which we gather that the $\lambda's$ belong to the interval $[0,1]$. Since they also add up to~1---recall that $\Tr(\hW)=1$---they have all the properties one expects from probabilities associated with mutually exclusive and jointly exhaustive events. A mixed state, therefore, defines probability distributions over probability distributions. It contains less information than a pure state. This additional lack of information may represent a lack of knowledge of relevant facts, but it may also be attributable to an additional lack of relevant facts---an objective indefiniteness over and above that associated with the individual algorithms $\ketbra{w_k}{w_k}$.

\section{How probabilities depend on measurement\\
outcomes}
\label{SecHowOutcomes}The trace rule tells us how to extract probabilities from a density operator~$\hW$. Our next order of business is to find~$\hW$. Suppose that $\hW_1$ is the density operator appropriate for assigning probabilities to the possible outcomes of whichever measurement we choose to make at the time~$t_1$. And suppose that a measurement~$M$ is made, and that the outcome represented by the projector~$\hP$ is obtained. What is the density operator~$\hW_2$ appropriate for assigning probabilities to the possible outcomes of whichever measurement is made next, at $t_2>t_1$?

As is customary in discussions of this kind, we concentrate on \textit{repeatable} measurements. If a system is subjected to two consecutive identical measurements, and if the second measurement invariable yields the same outcome as the first, we call these measurements repeatable. Accordingly, $\hW_2$ must satisfy the following conditions:
\ben
\item It is constructed out of $\hW_1$ and $\hP$.
\item It is self-adjoint.
\item $\Tr(\hW_2\hP)=1$.
\item $\Tr(\hW_2\hP_\perp)=0$, where $\hP_\perp$ is any possible outcome of~$M$ other than~$\hP$.
\een
The first and the fourth condition are satisfied by $\hW_1\hP$, $\hP\hW_1$, and $\hP\hW_1\hP$. Since $\hW_1\hP$ and $\hP\hW_1$ are not self-adjoint unless $\hW_1$ and $\hP$ commute, only $\hP\hW_1\hP$ also satisfies the second condition. To satisfy the third condition as well, all we have to do is divide by $\Tr(\hW_1\hP)=\Tr(\hP\hW_1)=\Tr(\hP\hW_1\hP)$. Thus,
\be
\hW_2=\frac{\hP\hW_1\hP}{\Tr(\hW_1\hP)}.
\ee
Now suppose that $M$ is a maximal test, and that $\hP=\ketbra ww$. Then
\be
\hW_2=\frac{\ketbra ww\hW_1\ketbra ww}{{\rm Tr}(\hW_1\ketbra ww)}=
\ket w\frac{\sandwich w{\hW_1}w}{\sandwich w{\hW_1}w}\bra w=\ketbra ww.
\ee
Lo and behold, if we update the density operator to take into account the outcome of a maximal test, it turns into the very projector representing this outcome. Observe that in this case $\hW_2$ is independent of~$\hW_1$.

\section{How probabilities depend on the times\\
of measurements}
The above definition of repeatability is nonstandard in that it requires $\hW_2$ to be independent of the time that passes between the two measurements. (The standard definition assumes that the second measurement is performed ``immediately'' after the first.) Let us relax this constraint and consider measurements that are \textit{verifiable}. A measurement~$M_1$, performed at the time~$t_1$, is verifiable if it is possible to confirm its outcome by a measurement~$M_2$ performed at the time $t_2>t_1$. If $M_1$ is repeatable, then it is verifiable by simply repeating it. Otherwise the verification requires (i)~the performance of a \textit{different} measurement and (ii)~the existence of a one-to-one correspondence between the possible outcomes of~$M_1$ and~$M_2$, which makes it possible to infer the actual outcome of~$M_1$ from the actual outcome of~$M_2$.

If the two measurements are maximal tests, then there are two ONBs, $\{\ket{a_i}\}$ and $\{\ket{b_k}\}$, such that the projectors $\{\ketbra{a_i}{a_i}\}$ represent the possible outcomes of $M_1$ and the projectors $\{\ketbra{b_k}{b_k}\}$ represent the possible outcomes of~$M_2$. The two bases are related by a unitary transformation: $\ket{b_k}=\hU\,\ket{a_k}$. In the absence of time-dependent boundary conditions, $\hU$~depends on the time difference $\Delta t_{21}=t_2-t_1$ but not, in addition, on $t_1$ or~$t_2$. If we imagine a third maximal test~$M_3$ with outcomes $\{\ketbra{c_i}{c_i}\}$, which verifies the outcome of $M_2$, we have $\ket{c_k}=\hU(\Delta t_{32})\,\ket{b_k}$ and thus
\be
\ket{c_k}=\hU(\Delta t_{32})\,\hU(\Delta t_{21})\,\ket{a_k}.
\ee
If we omit the second test, $M_3$ verifies $M_1$ directly (rather than indirectly, by verifying~$M_2$), and we have that
\be
\ket{c_k}=\hU(\Delta t_{32}+\Delta t_{21})\,\ket{a_k}.
\ee
Hence
\be
\hU(\Delta t_{32}+\Delta t_{21})=\hU(\Delta t_{32})\,\hU(\Delta t_{21}).
\label{EQUUU}
\ee
For every unitary operator $\hU$ there is a self-adjoint operator $\hA$ such that $\hU=e^{i\hA}$, the exponential being defined by its Taylor expansion.
Equation~(\ref{EQUUU}) is therefore equivalent to
\be
e^{i\hA(\Delta t_{32}+\Delta t_{21})}=e^{i\hA(\Delta t_{32})}\,e^{i\hA(\Delta t_{21})}.
\ee
But we also have that
\be
e^{i\hA(\Delta t_{32})}\,e^{i\hA(\Delta t_{21})}=e^{i[\hA(\Delta t_{32})+\hA(\Delta t_{21})]},
\ee
as one can check by expanding both sides in powers of $\hA$ and comparing coefficients. It follows that $\hA$ depends linearly on $\Delta t$:
\be
\hA(\Delta t_{32}+\Delta t_{21})=\hA(\Delta t_{32})+\hA(\Delta t_{21}).
\ee
This allows us to introduce a self-adjoint operator~$\hH$, known as the \textit{Hamilton operator} or simply the {\it Hamiltonian\/}, satisfying
\be
\hU(\Delta t)=e^{-(i/\hbar)\,\hH\,\Delta t}.
\ee
(The choice of the negative sign and the inclusion of the reduced Planck constant~$\hbar$ are of historical rather than physical interest.) For infinitesimal $dt$ we then have
\be
\hU(dt)=\hI-(i/\hbar)\,\hH\,dt.
\label{EQhUdt}
\ee
If the boundary conditions are time-dependent, then $\hU(dt)$ also depends on~$t$, which means that $\hH$ depends on~$t$. If $\hW(t_1)=\ketbra ww$, then $\hW(t_2)=\ketbra{w'}{w'}$ with $\ket{w'}=\hU(\Delta t_{21})\,\ket{w}$. For infinitesimal $\Delta t_{21}$ we therefore have that
\be
i\hbar\frac{d\,}{dt}\,\ket w=\hH\,\ket{w}.
\label{EQtevol}
\ee
Introducing (basis-dependent) matrix elements $H_{ik}=\sandwich{a_i}{\hH}{a_k}$ and vector components $w_i=\braket{a_i}{w}$, we obtain
\be
i\hbar\frac{dw_i}{dt}=\sum_k H_{ik}w_k,
\ee
and making the formal transition to continuous indices, we arrive at
\be
i\hbar\frac{dw(x,t)}{dt}=\int\!dx'\,H(x,x')\,w(x',t).
\label{EQHxx}
\ee
What should jump out at you right away is that the rate of change of $w$ at any point~$x$ depends on the \textit{simultaneous} value of w at any other point~$x'$. If we want to be consistent with the special theory of relativity, we need to introduce a delta distribution or define the Hamiltonian for the position representation via
\be
i\hbar\frac{dw(x,t)}{dt}=\hH\,w(x,t).
\label{EQtevolcont}
\ee
Because the integrand in Eq.~(\ref{EQHxx}) may depend on how $w(x)$ changes locally, across infinitesimal space intervals, $\hH$ may contain differential operators with respect to~$x$.

\section{Two fundamental Rules}
As mentioned in the Introduction, my quantum mechanics class begins with a statement of two simple premises and two simple rules. They make it possible to discuss key quantum phenomena in considerable depth and without much further mathematical ado.
\bd
\item[Premise a]Quantum mechanics provides us with algorithms for assigning probabilities to possible measurement outcomes on the basis of actual outcomes. Probabilities are calculated by summing over alternatives. \textit{Alternatives} are possible sequences of measurement outcomes. With each alternative is associated a complex number called \textit{amplitude}. (Observe that alternatives are defined in terms of measurement outcomes.)
\item[Premise b]To calculate the probability of a particular outcome of a measurement~$M_2$, given the actual outcome of an earlier measurement~$M_1$, choose a sequence of measurements that may be made in the meantime, and apply the appropriate Rule.
\item[Rule A]If the intermediate measurements are made (or if it is possible to infer from other measurements what their outcomes would have been if they had been made), first square the absolute values of the amplitudes associated with the alternatives and then add the results.
\item[Rule B]If the intermediate measurements are not made (and if it is not possible to infer from other measurements what their outcomes would have been), first add the amplitudes associated with the alternatives and then square the absolute value of the result.
\ed
Consider, for example, the two-slit experiment with electrons. $M_2$~is the detection of an electron somewhere at the backdrop. $M_1$~indicates the launch of the same electron in front of the slit plate. (If the electron gun~$G$ in front of the slit plate is the only source of free electrons, then the detection of an electron behind the slit plate also indicates the launch of an electron in front of the slit plate.) There are two alternatives, defined by the possible outcomes of a single intermediate measurement---``through the left slit~$L$'' and ``through the right slit~$R$.'' If Rule~A applies, the probability of detection at~$D$ is $p_A(D)=|A_L|^2+|A_R|^2$, where $A_L$ and~$A_R$ are the amplitudes associated with the alternatives. If Rule~B applies, we have $p_B(D)=|A_L+A_R|^2$ instead.

These amplitudes are not hard to find. Recall from Sec.~\ref{SecHowOutcomes} that a maximal test ``prepares from scratch.'' Or recall from Sec.~\ref{SecAxiom2} that $\Delta p\rightarrow\infty$ as $\Delta r\rightarrow0$. As a consequence, the probability $p(x|y)$ of finding a particle at~$x$, given that it was last ``seen'' at~$y$, is independent of whatever the particle did before its detection at~$y$. This implies that the propagation from $G$ to~$L$ and the propagation from $L$ to~$D$ are independent events. The probability for propagation from $G$ to~$D$ via~$L$ is therefore the product $p(D|L)\,p(L|G)$ of two probabilities. Accordingly, the corresponding amplitude is the product $\braket DL\braket LG$ of two amplitudes. Symmetry considerations then lead to the conclusion that for a free particle the amplitude $\braket xy$ can only depend on the distance~$d$ between $x$ and~$y$, and geometrical considerations lead to the conclusion that the absolute square of $\braket xy$ is inverse proportional to~$d^2$. The multiplicativity of successive propagators, finally, implies the additivity of their phases, whence once deduces that the phase of $\braket xy$ is proportional to~$d$. This is all the information that is needed to plot the distribution of ``hits'' at the backdrop predicted by Rules A and~B, respectively.

The validity of those Rules is readily demonstrated. Suppose, to begin with, that a maximal test at $t_1$ yields~$u$, and that we want to calculate the probability with which a maximal test at $t_2>t_1$ yields~$w$. Assume that at some intermediate time~$t$ another maximal test is made, and that its possible outcomes are the values $\{v_i|i=1,\dots,n\}$. Because a maximal test ``prepares from scratch,'' the joint probability $p(w,v_i|u)$ with which the intermediate and final tests yield~$v_i$ and~$w$, respectively, given the initial outcome~$u$, is the product of two probabilities: the probability $p(v_i|u)$ of~$v_i$ given~$u$, and the probability $p(w|v_i)$ of~$w$ given~$v_i$. By Born's rule (\ref{EQBorn}) this is 
\be
p(w,v_i|u)=|\braket w{v_i}\braket{v_i}u|^2.
\ee
The probability of $w$ given~$u$, \textit{irrespective} of the intermediate outcome, is therefore
\be
p_A(w|u)=\sum_i p(w,v_i|u)=\sum_i|\braket w{v_i}\braket{v_i}u|^2.
\ee
In other words, first take the absolute squares of the amplitudes $A_i=\braket w{v_i}\braket{v_i}u$ and then add the results.

If \textit{no} intermediate measurement is made, we have $p_B(w|u)=|\braket wu|^2$, and if we insert the identity operator $\hI=\sum_i\ketbra{v_i}{v_i}$, we have
\be
p_B(w|u)=\left|{\textstyle\sum_i}\braket w{v_i}\braket{v_i}u\right|^2.
\ee
In other words, first add the amplitudes $A_i$ and then take the absolute square of the result. The generalization to multiple intermediate measurements should be obvious.

\section{Composite systems}
\label{SecCompSyss}The two Rules stated in the previous section contain the clauses, ``if it is possible~/~not possible to infer from other measurements what their outcomes would have been~\dots'' To discover the rationale behind these cryptic phrases, we need to discuss composite systems. We begin with two independent systems, one associated with the vector~$\ket a$ from a vector space~${\cal V}_1$, the other associated with the vector~$\ket b$ from another vector space~${\cal V}_2$.  (``Independent'' here means that the probabilities of the possible outcomes of measurements that can be performed on one system do not depend on the actual outcomes of measurements to which the other system can be subjected.)

Considered as a pair $\ket{a,b}$, the two vectors belong to the direct product space ${\cal V}_1\otimes{\cal V}_2$. If the two systems carry identity tags of some sort, then the joint probability of finding the systems in possession of the respective properties $c$ and~$d$ is
\be
p(c,d|a,b)=p(c|a)\,p(d|b)=|\braket ca\braket db|^2.
\ee
Using the inner product on the direct product space, defined by $\braket{c,d}{a,b}=\braket ca\braket db$, we have
\be
p(c,d|a,b)=|\braket{c,d}{a,b}|^2.
\ee
Not every vector in ${\cal V}_1\otimes{\cal V}_2$ can be written in the form $\ket{a,b}$. A generic vector in this space has the form $\ket V=\sum_i\sum_k c_{ik}\ket{a_i,b_k}$, where $\{\ket{a_i}\}$ and $\{\ket{b_k}\}$ are ONBs in ${\cal V}_1$ and ${\cal V}_2$, respectively. It is, however, always possible to find a pair $\{\ket{A_i}\}$, $\{\ket{B_k}\}$ of ONBs such that $\ket V$ becomes a \textit{single} sum
\be
\ket V=\sum_i C_i\ket{A_i,B_i}
\label{EQbiorth}
\ee
of \textit{bi-orthogonal} terms~\cite{Peres5-3}.

Consider, for example, the ESW experiment mentioned in Section~\ref{SecAxiom2}. At the level of rigor we are working with, the state vector of the atom\,+\,cavity system, right after the atom's emergence from the slit plate, has the form
\be
\ket{V_{ESW}}=\frac1{\sqrt2}\Bigl[\ket{L,\lambda}+\ket{R,\rho}\Bigr],
\ee
where $\ket L$, $\ket R$, $\ket\lambda$, and $\ket\rho$ stand for the following measurement outcomes: atom went through the left slit, atom went through the right slit, photon is detected in the left cavity, photon is detected in the right cavity. The joint probability of obtaining the respective outcomes $\hP_a$ and $\hP_\gamma$ in measurements performed on the component systems equals
\bea
\label{EQesw}
&&\frac12\Bigl[\bra{L,\lambda}+\bra{R,\rho}\Bigr]\Bigl[\hP_a\otimes\hP_\gamma\Bigr]
\Bigl[\ket{L,\lambda}+\ket{R,\rho}\Bigr]\;=\\
&&\frac12\Bigl[\sandwich L{\hP_a}L\sandwich\lambda{\hP_\gamma}\lambda+
\sandwich R{\hP_a}R\sandwich\rho{\hP_\gamma}\rho+
\sandwich L{\hP_a}R\sandwich\lambda{\hP_\gamma}\rho+
\sandwich R{\hP_a}L\sandwich\rho{\hP_\gamma}\lambda\Bigr].\nonumber
\eea
Set $\hP_a$ equal to either $\ketbra LL$ or $\ketbra RR$ and $\hP_\gamma$ equal to either $\ketbra\lambda\lambda$ or $\ketbra\rho\rho$, and find that $p(L,\lambda)=p(R,\rho)=1/2$ and $p(R,\lambda)=p(L,\rho)=0$. The outcomes are perfectly correlated. It is therefore possible to infer from a measurement of the cavity containing the photon what we would have found if we had measured the slit taken by the atom. Thus Rule~A should apply. Let us see if it does. If the content of the microwave cavities remains unexamined, then $\hP_\gamma=\hI$, so that Eq.~(\ref{EQesw}) reduces to
\be
p(\hP_a)=\frac12\Bigl[\sandwich L{\hP_a}L+\sandwich R{\hP_a}R\Bigr].
\ee
The trace rule $p(\hP_a)=\Tr(\hW_a\hP_a)$ now tells us that
\be
\hW_a=\frac12\Bigl[\ketbra LL +\ketbra RR\Bigr],
\ee
a mixed state. The probability of finding the atom at~$D$ is thus
\be
\sandwich D{\hU\hW_a\hU^\dagger}D=\frac12\Bigl[\sandwich D\hU L\sandwich L{\hU^\dagger}D +\sandwich D\hU R\sandwich R{\hU^\dagger}D\Bigr],
\ee
where $\hU$ is the unitary operator that takes care of the time difference  between the (unperformed) intermediate measurement and the atom's detection at the backdrop. Finally we take into account that the atom was launched from a source~$S$. The factor $1/2$ is in fact nothing but the probability $|\sandwich L\hU S|^2=|\sandwich R\hU S|^2$ of either intermediate measurement outcome (given that the atom passes the slit plate). This allows us to write
\be
p\bigl(D|S)=\Bigl|\sandwich D\hU L\sandwich L\hU S\Bigr|^2+\Bigl|\sandwich D\hU R\sandwich R\hU S\Bigr|^2,
\ee
which is indeed what we get using Rule~A.

\section{Propagator for a free and stable scalar particle;\\
meaning of mass}
Imagine making $m$ intermediate position measurements at fixed intervals~$\Delta t$. The possible outcomes of each measurement are $n$~mutually disjoint regions~$R_k$, a partition of space. Under the conditions stipulated by Rule~$B$, the propagator $\sandwich B\hU A$ now equals the sum of amplitudes
\be
\sum_{k(1)=1}^n\cdots\sum_{k(m)=1}^n
\sandwich B\hU {R_{k(m)}}\cdots\sandwich{R_{k(2)}}\hU{R_{k(1)}}\sandwich{R_{k(1)}}\hU A.
\label{EQampsum}
\ee
It is not hard to see what happens in the double limit $\Delta t\rightarrow 0$ and $n\rightarrow\infty$: the alternatives (possible sequences of measurement outcomes) turn into continuous spacetime paths from $A$ to~$B$, and the sum becomes a path integral to which each spacetime path~$\cC$ contributes a complex number~$Z[\cC]$,
\be
\sandwich B\hU A=\int\DC Z[\cC].
\label{EQpathint}
\ee
(We use square brackets to indicate that $Z$ is a functional, which assigns numbers to \textit{functions}, in this case spacetime paths.) As it stands, the path integral $\int\DC$ is just the idea of an idea. Appropriate evalutations methods will have to be devised on a case-by-case basis. We proceed by picking an infinitesimal segment $d\cC$ of a path~$\cC$ leading from $A$ to~$B$. Let $t,x,y,z$ stand for the coordinates of its starting point and $t+dt,x+dx,y+dy,z+dz$ for those of its end point. If we choose inertial coordinates, meet the demands of the special theory of relativity, and consider a freely propagating, stable particle, we find that the only variable that $d\cC$ can depend on is the proper time interval
\be
ds=\sqrt{dt^2-(dx^2+dy^2+dz^2)/c^2}=dt\sqrt{1-v^2/c^2}.
\ee
Thus $Z(d\cC)=Z(ds)$. From the multiplicativity of successive propagators in Eq.~(\ref{EQampsum}) we deduce that
\be
\prod_{j=0}^m Z(ds_j)=Z\Bigl(\sum_{j=0}^m ds_j\Bigr)
\ee
and hence
\be
Z[\cC]=e^{z\,s[\cC]},
\ee
where $z$ is a constant. The functional $s[\cC]=\int_\cC ds$ is the time that passes on a clock as it travels from $A$ to~$B$ via~$\cC$. Integrating $\sandwich B\hU A$ (as a function of $\br_B$) over the whole of space, we obtain the probability of finding that a particle launched at~$\br_A$ at the time~$t_A$ still exists at the time~$t_B$. For a stable particle this equals~1:
\be
\int\!d^3r_B\,\Bigl|\sandwich{\br_B,t_B}\hU{\br_A,t_A}\Bigr|^2=
\int\!d^3r_B\left|\int\DC e^{z\,s[\cC]}\right|^2=1.
\label{qmch}
\ee
If you contemplate this equation with a calm heart and an open mind, you will notice that if the complex number $z=a+ib$ had a real part~$a\neq0$, then the expression in the middle would either blow up ($a>0$) or drop off ($a<0$) exponentially as a function of~$t_B$, due to the exponential $e^{a\,s[\cC]}$. Here at last we have a reason why the components of state vectors are complex numbers: $Z[\cC]$ has to be a phase factor $e^{ib\,s[\cC]}$.

Observe that the propagator for a free and stable scalar particle has a single ``degree of freedom'': the value of~$b$.  If proper time is measured in seconds, then $b$~is measured in radians per second. Nothings prevents us from thinking of $e^{ib\,s}$, with $s$ a proper-time parametrization of~$\cC$, as a clock carried by a particle traveling from $A$ to~$B$ via~$\cC$. (In doing so we visualize an aspect of the mathematical formalism of quantum mechanics, not something that actually happens.) It is customary
\bi
\item to insert a minus (so the clock actually turns clockwise!): $Z=e^{-ib\,s[\cC]}$,
\item to multiply by $2\pi$ (so we may think of $b$ as the rate at which the clock ``ticks''---the number of cycles it completes each second): $Z=e^{-i\,2\pi\,b\,s[\cC]}$,
\item to divide by Planck's constant~$h$ (so $b$ is measured in energy units and called the {\it rest energy\/} of the particle): $Z=e^{-i(2\pi/h)\,b\,s[\cC]}=e^{-(i/\hbar)\,b\,s[\cC]}$,
\item and to multiply by $c^2$ (so $b$ is measured in units of mass and called the particle's \textit{rest mass}): $Z=e^{-(i/\hbar)\,b\,c^2\,s[\cC]}$.
\ei
If two clocks, initially in the same place and ticking at the same rate, travel along different spacetime paths and then meet again, they still tick at the same rate. One second by one clock is still equal to one second by the other clock. Why? And why do different clocks indicate the same rate of time flow---one second of clock~A time per second of clock~B time---regardless of their constitution? Because the rates at which clocks tick (in the literal sense) ultimately depend on the rates at which free particles tick (in the figurative sense), and because the rate at which a free particle ticks (as measured in its rest frame) is determined by its mass, which is the same everywhere, anytime. (And why are two rods, initially in the same place and equally long, still equally long when they meet after having traveled along different spacetime paths? Because the meter is defined as the distance traveled by light in vacuum during a certain fraction of a second.) So much for the meaning of ``mass.''

\section{Energy and momentum}
In the real world, there is no such thing as a free particle. Some particles, however, are stable. For a stable scalar particle---no spin, no color, no flavor---$Z$~remains a phase factor: $Z[\cC]=e^{(i/\hbar)S[\cC]}$. We divided by~$\hbar$ to let the action $S[\cC]$ have its conventional units. The action $dS$ associated with an infinitesimal path segment $d\cC$ now depends on $dx,dy,dz,dt$ as well as $x,y,z,t$. The multiplicativity of successive propagators tells us that $dS(d\cC_1+d\cC_2)= dS(d\cC_1)+ dS(d\cC_2)$, where $d\cC_1$ and $d\cC_2$ are neighboring segments. It follows that $dS$ is homogeneous (of degree~1) in $dt$ and~$d\br$:
\be
dS(t,\br,\lambda\,dt,\lambda\,d\br)=\lambda\,dS(t,\br,dt,d\br).
\label{EQhomog}
\ee
Differentiating this equation with respect to $\lambda$, we obtain
\be
\frac{\d dS}{\d(\lambda\,dt)}dt+\frac{\d dS}{\d(\lambda\,d\br)}\cdot d\br=dS,
\ee
and setting $\lambda=1$, we arrive at
\be
\frac{\d dS}{\d dt}\,dt+\frac{\d dS}{\d d\br}\cdot d\br=dS.
\ee
Now suppose that the component $\d dS/\d dx$ of $\d dS/\d d\br$ is constant. This means that equal projections of $d\cC$ onto the x~axis make equal contributions to $dS$; they are physically equivalent, they represent equal distances. The same of course applies to $\d dS/\d dy$ and $\d dS/\d dz$. If $\d dS/\d d\br$ is constant, the spatial coordinates are homogeneous. But the vector quantity whose constancy is implied by the homogeneity of space is the momentum~$\bp$. By the same token, if $\d dS/\d dt$ is constant, then equal projections of $d\cC$ onto the time axis make equal contributions to $dS$; they represent equal time intervals. In other words, the time coordinate is homogeneous. But the quantity whose constancy is implied by the homogeneity of time is the energy~$E$. Thus,
\be
E=-\frac{\d dS}{\d dt},\qquad\bp=\frac{\d dS}{\d d\br}.
\ee
The reason why we define $E$ with a minus is that both $(c\,dt,dx,dy,dz)$ and $(E/c, p_x,p_y,p_z)$ are then 4-vectors, so their inner product
\be
-E\,dt+\bp\cdot d\br=dS
\ee
is a 4-scalar, as it should be. These definitions hold for particles that follow well-defined paths, \textit{including the alternatives contributing to} the path integral~(\ref{EQpathint}).

For a free particle we have
\be
dS=-mc^2\,ds=-mc^2\,dt\sqrt{1-v^2/c^2}.
\ee
The incorporation of effects on the motion of the particle---due to no matter what---requires a modification of~$dS$ that (i)~leaves it a 4-scalar and (ii)~preserves the validity of Eq.~(\ref{EQhomog}). The most straightforward such modification consists in adding a term that is not only homogeneous but also linear in $dt$ and~$d\br$:
\be
dS=-mc^2\,dt\sqrt{1-v^2/c^2}-qV(t,\br)\,dt+(q/c)\bA(t,\br)\cdot d\br.
\label{EQdSem}
\ee
We separated a particle-specific parameter $q$ called ``charge,'' and we inserted a $c$ so that $V$ and $\bA$ are measured in the same units.
All classical electromagnetic effects---to be precise, the Lorentz force law written in terms of the electromagnetic potentials $V$ and~$\bA$---can be derived from this simple modification of~$dS$. We are, however, headed in a different direction.

\section{Enter the wave function}
In the non-relativistic approximation to the real world, in which $v\ll c$, we expand the root in Eq.~(\ref{EQdSem}) to first order in $v^2/c^2$:
\be
\sqrt{1-v^2/c^2}\approx 1-{1\over2}{v^2\over c^2}.
\ee
In what follows we also ignore the effects that are represented by the vector potential~$\bA$, and we choose a particle of unit charge ($q=1$). We thus have
\be
S[\cC]=-mc^2(t_B-t_A)+\int_\cC dt\left[mv^2/2-V(t,\br)\right].
\ee
The path-independent term can be dropped, for all it contributes to the propagator $\sandwich B\hU A$ is an irrelevant overall phase factor. Hence
\be
\sandwich B\hU A=\int\DC e^{(i/\hbar)\int_\cC dt[mv^2/2-V(t,\br)]}.
\label{EQnrprop}
\ee
 It is customary to introduce a ``wave function'' $\psi(t,\br)$ such that
\be
\psi(\br_B,t_B)=\int\!d^3r_A\,\sandwich{\br_B,t_B}\hU{\br_A,t_A}\,\psi(\br_A,t_A).
\label{EQwfprop}
\ee
Feynman~\cite{FH,Derbes} has shown how to get from here to the Schr\"odinger equation
\be
i\hbar{\d\psi\over\d t}=-{\hbar^2\over2m}\left({\d^2\psi\over\d x^2}+
{\d^2\psi\over\d y^2}+{\d^2\psi\over\d z^2}\right)+V\psi.
\label{EQschr}
\ee
Compare this with Eq.~(\ref{EQtevolcont}).

\section{Discussion}
Undoubtedly the most effective way of introducing quantum mechanics is the axiomatic approach. Philosophically, however, this approach has its dangers. Axioms are supposed to be clear and compelling. The standard axioms of quantum mechanics lack both desiderata. Worse, given the lack of a convincing physical motivation, students---but not only students---tend to accept them as ultimate encapsulations of the way things are.
\ben
\item The first standard axiom typically tells us that the state of a system~$S$ is (or is represented by) a normalized element~$\ket v$ of a Hilbert space~$\cH_S$. (Weinberg~\cite{WeinbergQFT} is nearer the mark when he represents the state of~$S$ by a \textit{ray} in~$\cH_S$.)
\item The next axiom (or couple of axioms) usually states that observables are (or are represented by) self-adjoint linear operators on~$\cH_S$, and that the possible outcomes of a measurement of an observable~$\hO$ are the eigenvalues of~$\hO$.
\item Then comes an axiom (or a couple of axioms) concerning the time evolution of states. Between measurements (if not always), states are said to evolve according to unitary transformations, whereas at the time of a measurement they are said to evolve (or appear to evolve) as stipulated by the projection postulate. That is, if $\hO$ is measured, the subsequent state of~$S$ is the eigenvector corresponding to the outcome, regardless of the previous state of~$S$.
\item A further axiom stipulates that the states of composite systems are (or are represented by) vectors in the tensor product of the Hilbert spaces of the component systems.
\item Finally there is an axiom (or a couple of axioms) concerning probabilities. If $S$ is in the state~$\ket v$ and we do an experiment to see if it has the property $\ketbra ww$, then the probability~$p$ of a positive outcome is given by the Born rule $p=|\braket wv|^2$. Furthermore, the expectation value of an observable~$\hO$ in the state~$\ket v$ is $\sandwich v\hO v$.
\een
There is much here that is perplexing if not simply wrong. To begin with, what is meant by saying that the state of a system is (or is represented by) a normalized vector (or else a ray) in a Hilbert space? The main reason why this question seems all but unanswerable is that probabilities are introduced almost as an afterthought. It ought to be stated at the outset that \textit{the mathematical formalism of quantum mechanics provides us with algorithms for calculating the probabilities of measurement outcomes}. If both the phase space formalism of classical physics and the Hilbert space formalism of quantum physics are understood as tools for calculating the probabilities of measurement outcomes, the transition from a 0-dimensional point in a phase space to a 1-dimensional ray in a Hilbert space is readily understood as a straightforward way of making room for nontrivial probabilities. Because the probabilities assigned by the points of a phase space are trivial, the classical formalism admits of an alternative interpretation: we may think of states as collections of possessed properties. Because the probabilities assigned by the rays of a Hilbert space are nontrivial, the quantum formalism does not admit of such an interpretation.

Saying that the state of a quantum system is (or is represented by) a vector or a ray in a Hilbert space is therefore highly ambiguous if not seriously misleading. \textit{At} the time of a measurement, a system (or observable) has the property (or value) that is indicated by the outcome. It is true that measurement outcomes are represented by projectors or the corresponding subspaces, but \textit{solely} for the purpose of calculating probabilities. Thus if $\ketbra vv$ represents the outcome of a maximal test \textit{for the purpose of assigning probabilities to the possible outcomes of whichever measurement is subsequently made}, and if $\ketbra ww$ represents a possible outcome of this measurement \textit{for the purpose of assigning to it a probability}, then the probability of this outcome is $\Tr(\ket v\braket vw\bra w)=|\braket wv|^2$. If the Hamiltonian is not zero, a unitary operator has to be sandwiched between $\bra w$ and~$\ket v$.

We ought to have the honesty to admit that any statement about a quantum system \textit{between measurements} is ``not even wrong'' (Pauli's famous phrase), inasmuch as such a statement is by definition neither verifiable nor falsifiable experimentally. This bears on the third axiom (or couple of axioms), according to which quantum states evolve (or appear to evolve) unitarily between measurements, which implies that they ``collapse'' (or appear to collapse) at the time of a measurement.

Stories of this kind are based on a misunderstanding of the time dependence of quantum states. $\ket{v(t)}$ is not an instantaneous state of affairs that obtains at the time~$t$. It is not something that evolves, nor do Eqs. (\ref{EQtevol}), (\ref{EQtevolcont}), and (\ref{EQschr}) describe the evolution of a state of affairs of any kind. As Asher Peres pointedly said, ``there is no interpolating wave function giving the `state of the system' between measurements''~\cite{PeresSV}. $\ket{v(t)}$~is an algorithm for assigning probabilities to the possible outcomes of a measurement performed at the time~$t$. The parameter~$t$ refers to the time \textit{of this measurement}.

Again, what is meant by saying that observables are (or are represented by) self-adjoint operators? We are left in the dark until we get to the last axiom (or couple of axioms), at which point we learn that the expectation value of an observable~$\hO$ in the state~$\ket v$ is $\sandwich v\hO v$. This expectation value (whose probability, incidentally, can be~0) is nothing but the mean value $\sum_k O_k\,|\braket{O_k}v|^2$. So what does~$\hO$ \textit{qua operator} have to do with it? If we \textit{define} a self-adjoint operator $\hO=\sum_k\,O_k\,\ketbra{O_k}{O_k}$, where the projectors $\ketbra{O_k}{O_k}$ represent the possible outcomes~$O_k$ (for the purpose of assigning probabilities to these outcomes), then the expectation value can be written as $\sandwich v{\hO}v$. Because $\hO$ is self-adjoint, its eigenvalues $O_k$ are real, as the possible outcomes of a measurement should be. That is all there is to the mysterious claim that observables are (represented by) self-adjoint operators.

And finally, why would the states of composite systems be (represented by) vectors in the tensor product of the Hilbert spaces of the component systems? Once again the answer is almost self-evident (recall Sec.~\ref{SecCompSyss}) if quantum states are seen for what they are---tools for assigning nontrivial probabilities to the possible outcomes of measurements.

There remains the question of why the fundamental theoretical framework of contemporary physics is  a collection of tools for calculating the (in general) nontrivial probabilities of measurement outcomes. One thing is clear: a fundamental theory cannot be explained with the help of a more fundamental theory; ``fundamental'' has no comparative. There will always be a fundamental theory, and there will always be the mystery of its origin. If at all a fundamental theory can be explained, it is in weakly teleological terms. (Strongly teleological claims are not part of physics.) Where quantum mechanics is concerned, such an explanation can be given by pointing to the existence of objects that (i)~``occupy space,'' (ii)~are composed of a finite number of objects that do not occupy space, and (iii)~are stable (Sec.~\ref{SECwhyprobas}). The existence of such objects is made possible by the fuzziness of their internal relative positions and momenta, and the proper way of dealing with a fuzzy observable~$O$ is to assign nontrivial probabilities to the possible outcomes of a measurement of~$O$. (Let us not forget that it was the instability of Rutherford's model of the atom that led Bohr to postulate the quantization of angular momentum, arguably the single most seminal event in the history of quantum mechanics.)

This leaves me, as a teacher, with the task of conveying an intuitive understanding, or ``feel,'' for the fuzziness of the quantum world. The performance of this task is inexorably intertwined with (i)~an analysis of the quantum-mechanical probability assignments in a variety of experimental contexts and (ii)~a discussion of their possible ontological implications. This is a complex task, which is best undertaken in a separate article. Meanwhile interested readers will find these topics discussed in Refs.~\cite{Mohrhoff05} and \cite{Mohrhoff04}. What I attempted to demonstrate in the present article is how readily quantum mechanics is understood---how much sense it makes, how \textit{obvious} it becomes---if it is treated as a general probability algorithm, as compared to the impenetrability of its standard axioms, not to speak of the pseudo-problems that arise if quantum states are taken for evolving states of affairs of any kind.

\pagebreak


\begin{thebibliography}{00}

\bibitem{Feynmanteach}
R P Feynman, R B Leighton, and M Sands, \textit{The Feynman Lectures in Physics}, Vol.~1 (Addison-Wesley, Reading, 1963) p. 1--2

\bibitem{QED}
R P Feynman, \textit{QED: The Strange Theory of Light and Matter} (Princeton University Press, Princeton, 1985) p.~12

\bibitem{Mackey}
G W Mackey, \textit{The mathematical Foundations of Quantum Mechanics} (Benjamin, New York, 1963)

\bibitem{Varadarajan}
V Varadarajan, \textit{Geometry of Quantum Theory} (Springer, New York, 1968)

\bibitem{Ludwig}
G Ludwig, \textit{Foundations of Quantum Mechanics}, Vols. I and II (Springer, New York, 1983, 1985)

\bibitem{Pitowsky89}
I Pitowsky, \textit{Quantum Probability--Quantum Logic}, Lecture Notes in Physics 321 (Springer, Berlin, 1989)

\bibitem{Lieb}
E H Lieb, \textit{Rev. Mod. Phys.} \textbf{48}, 553 (1976)

\bibitem{Bell}
J S Bell, in \textit{62 Years of Uncertainty}, edited by A~I Miller (Plenum, New York, 1990), p.~17

\bibitem{Mermin}
N D Mermin, \textit{Am. J. Phys.} \textbf{66}, 753 (1998)

\bibitem{Primas}
H Primas, \textit{Mind and Matter} \textbf{1}, 81 (2003)

\bibitem{Jauch}
J M Jauch, \textit{Foundations of Quantum Mechanics} (Addison-Wesley, Reading, 1968), pp. 92--94, 132
 
\bibitem{Crease}
R P Crease, \textit{Physics World}, 19 (September 2002)

\bibitem{FLS}
R P Feynman, R B Leighton, and M Sands, \textit{The Feynman Lectures in Physics}, Vol.~3 (Addison-Wesley, Reading, 1965), Secs. 1--1 to 1--6

\bibitem{GZ}
G Greenstein and A Zajonc, \textit{The Quantum Challenge: Modern Research on the Foundations of Quantum Mechanics} (Jones and Bartlett, Sudbury, 1997), Chap.~1

\bibitem{Jonnson}
C J\"onnson, \textit{Am. J. Phys.} \textbf{42}, 4 (1974)

\bibitem{Tonomura}
A Tonomura, J Endo, T Matsuda, T Kawasaki, and H Ezawa, \textit{Am. J. Phys.} \textbf{57}, 117 (1989)

\bibitem{SEW}
M O Scully, B-G Englert, and H Walther, \textit{Nature} \textbf{ 351}, 111 (1991)

\bibitem{ESW}
B-G Englert, M O Scully, and H Walther, \textit{Sci. Am.} \textbf{271}, 56 (1994)
 
\bibitem{Mohrhoff99} 
U Mohrhoff, \textit{Am. J. Phys.} \textbf{67}, 330 (1999)

\bibitem{Mohrhoff01}
U Mohrhoff, \textit{Am. J. Phys.} \textbf{69}, 864 (2001)

\bibitem{PeresCh7}
A Peres, \textit{Quantum Theory: Concepts and Methods} (Kluwer, Dordrecht, 1995), Chap.~7

\bibitem{Mermin93}
N D Mermin, \textit{Rev. Mod. Phys.} \textbf{65}, 803 (1993)

\bibitem{Clifton}
R Clifton, \textit{Am. J. Phys.} \textbf{61}, 443 (1974)

\bibitem{Peres190}
A Peres, \textit{Quantum Theory: Concepts and Methods} (Kluwer, Dordrecht 1995), p.~190

\bibitem{Gleason}
A M Gleason, \textit{J. Math. Mech.} \textbf{6}, 885 (1957)

\bibitem{Pitowsky98}
I Pitowsky, \textit{J. Math. Phys.} \textbf{39}, 218 (1998)

\bibitem{Cookeetal}
R Cooke, M Keane, and W Moran, \textit{Math. Proc. Camb. Phil. Soc.} \textbf{98}, 117 (1985)

\bibitem{Fuchs2001}
C A Fuchs, in \textit{Decoherence and its Implications in Quantum Computation and Information Transfer} edited by A Gonis and P~E~A Turchi (IOS Press, Amsterdam, 2001), pp. 38--82

\bibitem{Busch}
P Busch (2003), \textit{Phys. Rev. Lett.} \textbf{91}, 120403

\bibitem{Cavesetal}
C Caves, C A Fuchs, K Manne, and J M Rennes, \textit{Found. Phys.} \textbf{34}, 193 (2004)

\bibitem{Peres9-5}
A Peres, \textit{Quantum Theory: Concepts and Methods} (Kluwer, Dordrecht, 1995), Sec.~9--5

\bibitem{Peres5-3}
A Peres, \textit{Quantum Theory: Concepts and Methods} (Kluwer, Dordrecht, 1995), Sec.~5--3

\bibitem{FH}
R P Feynman and A R Hibbs, \textit{Quantum Mechanics and Path Integrals} (McGraw-Hill, New York, 1965)

\bibitem{Derbes}
D Derbes, \textit{Am. J. Phys.} \textbf{64}, 881 (1996)

\bibitem{WeinbergQFT}
S Weinberg, \textit{The Quantum Theory of Fields}, Vol. 1 (Cambridge University Press, Cambridge,  1996), pp. 49--50

\bibitem{PeresSV}
A Peres, \textit{Am. J. Phys.} \textbf{52}, 644 (1984)

\bibitem{Mohrhoff05}
U Mohrhoff, \textit{Pramana--J. Phys.} \textbf{64}, 171 (2005)

\bibitem{Mohrhoff04}
U Mohrhoff, \textit{Int. J. Q. Inf.} \textbf{2}, 201 (2004)

\end{thebibliography}
\end{document}